\documentclass[aip, apl, amsmath,amssymb, preprint, numerical]{revtex4-1}
\usepackage{graphicx}
\usepackage{dcolumn}
\usepackage{bm}
\usepackage{xcolor}
\usepackage{siunitx}
\definecolor{red}{HTML}{ff474c}
\definecolor{blue}{HTML}{0343df}
\definecolor{grey}{HTML}{363737}
\definecolor{orange}{HTML}{f97306}

\begin{document}

\preprint{APL}
\title[]{Noise in stimulated Raman scattering measurement: From basics to practice.}
\author{X. Audier}
\affiliation{Aix Marseille Univ, CNRS, Centrale Marseille, Institut Fresnel, Marseille, France}
\author{S. Heuke}
\affiliation{Aix Marseille Univ, CNRS, Centrale Marseille, Institut Fresnel, Marseille, France}
\author{P. Volz}
\affiliation{APE Angewandte Physik \& Elektronik GmbH, Berlin, Germany}
\author{I. Rimke}
\affiliation{APE Angewandte Physik \& Elektronik GmbH, Berlin, Germany}
\author{H. Rigneault}
\email{herve.rigneault@fresnel.fr}
\affiliation{Aix Marseille Univ, CNRS, Centrale Marseille, Institut Fresnel, Marseille, France}
\date{\today}

\begin{abstract}
We revisit laser intensity noise in the context of stimulated Raman scattering (SRS), which has recently proved to be a key technique to provide label free images of chemical bonds in biological and medical samples.
Contrary to most microscopy techniques, which detect a weak photon flux resulting from light matter interactions, SRS is a pump-probe scheme that works in the high flux regime and happens as a weak modulation ($10^{-4}-10^{-6}$) in a strong laser field.
As a result, laser noise is a key issue in SRS detection.
This practical tutorial provides the experimentalists with the tools required to assess the amount of noise and the ultimate SRS detection limit in a conventional lock-in-based SRS system.
We first define the quantities that are relevant when discussing intensity noise, and illustrate them through a conventional model of light detection by a photodiode.
Stimulated Raman Scattering is then introduced in its lock-in-based implementation, and the model presented is adapted in this particular case.
The power spectral density (PSD), relative intensity noise (RIN), signal to noise ratio (SNR), and sensitivity of the system are derived and discussed.
Two complementary methods are presented that allow measurement of the RIN and assessment of the performance of a SRS system.
Such measurements are illustrated on two commercial laser systems.
Finally, the consequences of noise in SRS are discussed, and future developments are suggested.
The presentation is made simple enough for under-graduated, graduated students, and newcomers in the field of stimulated Raman, and more generally in pump-probe based schemes. 
\end{abstract}
\maketitle

\section{\label{sec:intro}Introduction}
Over the last two decades, coherent Raman imaging has evolved as a mature, label-free, imaging technique with numerous applications in biology and medicine~\cite{Cheng2015}.
The seminal work of Zumbusch in 1999 revived coherent anti-Stokes Raman scattering (CARS)~\cite{Avenger2009} as a vibrational microscopic imaging modality~\cite{Zumbusch1999}.
Since then, the coherent Raman imaging field experienced a second revolution in 2008 when stimulated Raman scattering (SRS)~\cite{Bloembergen1967} was also demonstrated as a powerful vibrational imaging scheme~\cite{Freudiger2008, Nandakumar2009}. 
Contrary to CARS, SRS is free of nonresonant background and scales linearly with the molecular concentration~\cite{Rigneault2018}.
These key features initiated the development of SRS imaging technologies~\cite{He2017, Berto2017, Alshaykh2017, Liao2016, Liao2015, Liao2015b, Berto2014, Freudiger2014a, Heuke2018b, Heuke2018a} and facilitated their successful applications in biology~\cite{Lu2015a, Fu2017, Ozeki2012, Wakisaka2016}, chemistry~\cite{Wei2017a, Winterhalder2015} and medicine~\cite{Ji2015a, Orringer2017, Sarri2019, Cicerone2018} as a quantitative and label free chemical imaging modality.

Contrary to CARS, which detects faint generated photons at specific wavelengths, SRS is a pump-probe scheme that works in the high photon flux regime.
It manifests itself as a weak modulation ($10^{-4}- 10^{-6}$) that is transferred from an amplitude modulated (AM) laser on an unmodulated (probe) laser beam~\cite{Freudiger2008}.

Because the modulation transferred to the probe beam is weak, the laser noise and the detection of electronic noise are key components to achieve the ultimate SRS detection level~\cite{Ozeki2009, Ozeki2010a}. 
For instance, performing SRS with noisy fiber lasers requires the development of specific balanced detection schemes~\cite{Freudiger2014a, Crisafi2017, Laptenock2019}, which are at best \SI{3}{\dB} above the shot noise limit. 
Although they are key to SRS, it appears that considerations on laser and detection noise in SRS imaging systems are still lagging behind when compared to the demonstrated technological and application advances.
For instance, it is often not clear in published papers how the reported SRS detection ability compares to the shot noise limit.

This paper is intended to provide the SRS experimentalist with theoretical basics and most importantly a reliable experimental method to characterize the noise of an SRS system as compared to the ultimate shot noise.
More generally, our presentation applies also to any pump probe spectroscopy scheme.
Although no fundamental breakthrough is presented here, it is our understanding that the required background to master noise physical description and measurement is often not available to physical chemists that are building and using SRS systems.
The scope of this paper is to provide a tutorial on noise found in SRS systems accessible for under-graduated, graduated students, and newcomers in the field. 

We start by defining noise in the context of light intensity detection, and present a conventional model used to describe such noise.
The important quantities - power spectral density (PSD), relative intensity noise (RIN), and signal to noise ratio (SNR) - are calculated for this model.
We then present a typical lock-in-based SRS measurement system and describe how the previous model changes in this context.
The SNR and sensitivity of such SRS system are discussed and linked to the laser RIN.
Two complementary measurements of the RIN are presented, that can be used to characterize and optimize the performance of a SRS system.
These measurements are illustrated using two commercial laser systems commonly used for SRS imaging.
We finally discuss the implications of such measurements and additional means to further increase the signal to noise ratio in SRS in both current systems and in future technological developments. The paper is complemented with a supplementary information section that is intended to be used as a practical handbook to perform RIN noise measurement of a SRS laser system.
For a comprehensive introduction to coherent Raman and SRS we invite the reader to refer to the tutorials~\cite{Rigneault2018}.  

\section{Definitions and laser intensity noise model\label{sec:model1}}
Although various implementations of SRS have been developed, they mostly share the same working principle, which is the detection of small intensity variations in an intense laser beam.
In this section, a standard model of laser intensity measurement is presented, along with the definitions of the power spectral density (PSD), relative intensity noise (RIN), and signal to noise ratio (SNR).
The RIN of a laser is calculated for this model, to illustrate the contribution of classical intensity noise, quantum fluctuations, and electrical noise.



\subsection{Model}
The optical intensity of a laser is modeled by a time-varying photon rate $\mathcal{I}_{opt}(t)$.
In this tutorial, the term \emph{optical intensity} is preferred to optical power in order to avoid confusion when discussing electrical power later.
The laser is assumed monochromatic for simplification and is detected with a photodiode.

The electrical current $I(t)$ at the output of this photodiode is modeled using a semi-classical approach, consistent with the detection of coherent states of light.
Similar work has been done by Quinlan et al.~\cite{Quinlan2013} and can be found in textbooks~\cite{saleh1991}.

The detection of a single photon at time $t=0$ produces an electrical current $h(t)$ at the output of the photodiode.
The temporal spread of $h(t)$ provides the bandwidth of the photodiode, and the area under the curve $h(t)$ is equal to a single electric charge $q$:
\begin{equation}
\label{eq:ht}
\int_{-\infty}^{+\infty} h(t) \mathrm{d}t = q 
\end{equation}
In this work, the contribution of the photodiode will often be simplified by assuming that its bandwidth is greater than all electronic frequencies of interest.
In the time domain, $h(t)$ will be approximated by a Dirac distribution.
In the frequency domain, $\hat{h}(f)$ will therefore be assumed to be flat $\hat{h}(f) \approx \hat{h}(0) = q$.

Because SRS is a nonlinear optical process that requires short optical pulses~\cite{Rigneault2018}, it is necessary to model pulsed lasers, consisting of intense photon bursts occurring at time scales of femtoseconds to picoseconds.
Since the pulse duration is much shorter than the electronic bandwidth of most measurement devices, all the photons in a single pulse are considered to arrive simultaneously.
For pulsed laser light (derived from mode-locked oscillators, optical parametric oscillators and amplifiers)  the successive pulses are numbered with the integers $k \in \mathbb{Z}$.
The time of arrival of pulse $k$ is $t_k = k \times T_r$, where $T_r$ is the period of the oscillator, the inverse of the repetition rate $f_r$.
The number of electrons generated by the optical pulse $k$ is a random variable $X_k$ which follows a Poisson distribution of mean $N(t_k)$.
The average number of electron generated per pulse is linked to the optical intensity $\mathcal{I}_{opt}(t)$ via the photon-to-electron conversion equation:
\begin{equation}
N(t_k) = \frac{\eta}{h \nu} \int_{pulse \, k} \mathcal{I}_{opt}(t) \mathrm{d}t
\label{eq:p(t)}
\end{equation}
Where $\eta$ is the detector quantum efficiency, $\nu$ the optical frequency, and $h$ the Plank constant.
In addition to the photo-detection events, the detector current noise is modeled with a stochastic current $\epsilon(t)$. This current noise encompasses the detector dark current, the Johnson-Nyquist noise of the load resistor, and any other sources of electrical noise that is generated at the output of the photodiode independently of the optical intensity.
With this model, the electrical current $I(t)$ can be expressed by:
\begin{equation}
\label{eq:photocurrentmodel}
I(t) = \Big[ \sum_k X_k \delta(t - t_k) \Big] \otimes h(t) + \epsilon(t)
\end{equation}
where $\otimes$ is the convolution operator.
The electrical noise $\epsilon(t)$ is considered to be independent from the $\{X_k\}$, and for simplification it will be assumed that the electrical noise has zero mean: $\langle \epsilon(t) \rangle = 0$.
Here, $\langle \cdot \rangle $ stands for the ensemble average, meaning the average over all possible realizations of the measurement given the exact same system in the exact same state.
This is analogous to having a large number of identical systems all performing the same measurement.
For such ensemble of systems, each will record different values for $X_k$, but these values will be distributed in a Poisson law of average $N(t_k)$.
The laser intensity fluctuations over time have therefore two origins. The first one is the fluctuation of $N(t_k)$, which arises from generation of optical pulses that are not perfectly identical, and is referred to as classical noise here. 
The second is the randomness on the measurement and is called shot noise. It causes a fluctuation of the value $X_k$ even when pulses would be perfectly engineered ($N(t_k)$ = constant).
From the model described by equation~\ref{eq:photocurrentmodel}, one can compute the power contribution of the DC component and frequency components of the electrical current $I(t)$ through a load resistor R.

\subsection{Definitions}
Let us consider a quantity $A$ that is a function of time. In the following, definitions are made using the general quantity $A$, although $A$ will be replaced by photodiode current $I$ for calculations. Similar derivations could be done using voltages but the discussion, in particular in terms of power, signal, and noise, would be identical.
\subsubsection{DC power}
In order to study the contribution of different frequencies to the electrical current, the finite-time Fourier transform~\cite{Gardner1986a} of a quantity $A(t)$ is defined as:
\begin{equation}
\mathcal{F}_T\{A(t)\}(f) \equiv \hat{A}_T(f) \equiv \int^{T/2}_{-T/2} A(t)e^{-2 i \pi f t}\mathrm{d}t \label{eq:FT}
\end{equation}
The time average of $A(t)$ is defined as:
\begin{equation}
\label{eq:Iavg}
A_{avg} \equiv \lim_{T\rightarrow \infty} \frac{1}{T} \langle \hat{A}_T(0) \rangle
\end{equation}
The total power of quantity $A$ is defined as the time average of $A^2$:
\begin{equation}
\label{eq:PtotA}
P_{tot}[A] \equiv (A^2)_{avg}
\end{equation}
Using the definition of the time variance $Var[A] \equiv (A^2)_{avg} - A_{avg}^2$, the total power can be split into two components:
\begin{equation}
P_{tot}[A] = A_{avg}^2 + \mathrm{Var}[A]
\end{equation}
These two terms have two distinct origins, the first is the DC power, the second is the power carried by the fluctuations of $A$.
Typically when measuring the value of $A$, the DC power corresponds to the signal power, and the power of the fluctuations are referred to as noise power $P_{Noise}$.

Note that "power" as defined here differs by a constant from the usual electrical power expressed in Watts.
The power defined by equation~\ref{eq:PtotA} has the dimension of $[A]^2$, where $[A]$ is the unit of $A$.
To obtain an electrical power in $W$, one needs to multiply by the appropriate factor, such as $R$ if $A$ is a current or $1/R$ if $A$ is a voltage.
For instance, the DC power and noise power of a voltage $V$ generating a current $I$ passing through a load resistor R are:
\begin{align}
P_{DC} &= V_{avg}^2 / R = R \, I_{avg}^2 \label{eq:Pavg} \\
P_{Noise} &= \mathrm{Var}[V] / R = R \, \mathrm{Var}[I] \label{eq:PNoise}
\end{align}
The noise power expressed through the variance does not provide information on the frequency at which the fluctuations of $A$ are happening.

\subsubsection{Power Spectral Density (PSD)}
To study how the different frequencies contribute to $\mathrm{Var}[A]$, and therefore to the noise power, the double-sided (positive and negative frequencies) power spectral density of a quantity $A(t)$ is defined as:
\begin{equation}
\label{eq:SA}
S_A(f) \equiv \lim_{T\rightarrow \infty} \frac{1}{T} \langle \vert\hat{A}_T(f)\vert^2 \rangle
\end{equation}
Because $A(t)$ is usually a real quantity, $S_A$ is an even function of frequency.
For this reason, only the positive frequencies are usually considered and the single-sided power spectral density of a quantity $A(t)$ is defined as:
\begin{equation}
\label{eq:SSA}
S^+_A(f) \equiv 2 S_A(f); f>0
\end{equation}
As mentioned above, the name "power spectral density" is ambiguous, as it usually does not have the dimension of a power density, but rather of $[A]^2 / Hz$, where $[A]$ is the unit of $A$.
The PSD measures the amount of electrical power (variance) in the signal per unit of bandwidth.
For instance, the amount of power coming from the frequency range with width $\Delta f$ centered around $f_0$ is given by:
\begin{equation}
P_{\Delta f}(f_0) = \int_{f_0 \pm \Delta f / 2} R \, S^+_I(f) \mathrm{d}f
\label{eq:PDf}
\end{equation}

\subsubsection{Relative Intensity Noise (RIN)}
The power spectral density divided by $A_{avg}^2$ gives the relative intensity noise (RIN) of quantity $A$:
\begin{equation}
\label{eq:RINA}
\mathcal{RIN}_A(f, A_{avg}) \equiv \frac{S^+_A(f)}{A_{avg}^2}
\end{equation}
The RIN is expressed in "per unit bandwidth" $1 / Hz$ and quantifies the relative contribution of each spectral component to the total signal power.

\subsubsection{Signal to Noise Ratio (SNR)}
When measuring DC value of the quantity $A(t)$, the signal power is given by the DC electrical power $A_{avg}^2$.
The power of the noise is given by equation~\ref{eq:PDf} where the integral covers the bandwidth $\Delta f$ of the measurement system (typically $f \in [0, \Delta f]$, to measure the DC component).
The signal to noise ratio can therefore be expressed as:
\begin{equation}
\mathcal{SNR} = \frac{A_{avg}^2}{\int_{0}^{\Delta f} \, S^+_A(f) \mathrm{d}f}
\end{equation}
As a result, the SNR for a measurement of the average value of $A$ is inversely related to the RIN:
\begin{equation}
\mathcal{SNR}^{-1} = \int_{0}^{\Delta f} \mathcal{RIN}_A(f, A_{avg}) \mathrm{d}f
\label{eq:SNRRIN}
\end{equation}

\subsection{Calculations}
\subsubsection{Average and relative intensity}
Under the assumption that the electrical noise has a null time average, the time average current $I_{avg}$ can be computed (equation~\ref{eq:ht},~\ref{eq:photocurrentmodel},~\ref{eq:FT}, and~\ref{eq:Iavg}) in terms of the time average number of electrons generated $N_{avg}$:
\begin{align}
\label{eq:Iavg2}
I_{avg} &= \lim_{T\rightarrow \infty} \frac{1}{T} \langle \int^{T/2}_{-T/2} \Big( \Big[ \sum_k X_k \delta(t - t_k) \Big] \otimes h(t) + \epsilon(t) \Big) e^{-2 i \pi f t}\mathrm{d}t \rangle \vert_{f=0} \nonumber\\
&= \lim_{T\rightarrow \infty} \frac{1}{T}  \int^{T/2}_{-T/2} \sum_k \langle X_k \rangle h(t - t_k)\mathrm{d}t \nonumber \\
&= \lim_{T\rightarrow \infty} \frac{1}{T}  \int^{T/2}_{-T/2} \sum_k N(t_k) h(t - t_k)\mathrm{d}t \nonumber \\
&= \lim_{T\rightarrow \infty} \frac{1}{T} \sum_{t_k \in [-T/2, T/2]} q N(t_k) \nonumber \\
&= q N_{avg}
\end{align}
The average photocurrent $I_{avg}$ delivered by the photodiode can either be measured directly with an oscilloscope, or derived from the laser average intensity $\mathcal{I}_{opt, avg}$ using the following relationship:
\begin{equation}
I_{avg} = \frac{q \eta}{h \nu} \mathcal{I}_{opt, avg} \label{eq:Ioptavg}
\end{equation}
The relative intensity $\alpha(t)$ is defined as the expected current divided by the average current:
\begin{equation}
\alpha(t) \equiv \frac{\langle I(t) \rangle}{I_{avg}} = \frac{N(t)}{N_{avg}}
\label{eq:alphat}
\end{equation}

\subsubsection{Relative Intensity Noise\label{sec:RINcalc1}}
The photocurrent power spectral density $S_I(f)$ can be computed by considering the photodetection events statistically independent:
\begin{equation}
\langle X_k X_l \rangle = N(t_k) N(t_l) + N(t_k) \delta_{k, l} \label{eq:PkPkk}
\end{equation}
The PSD from the model defined here reads (Annex A):
\begin{equation}
S_I(f) = S_\epsilon(f) + \vert \hat{h}(f) \vert^2 \Big[ N_{avg} + S_N(f) \Big]
\label{eq:PSD}
\end{equation}
where $\hat{h}(f)$, the Fourier transform of $h(t)$, is the spectral response of the detector.
For frequencies within the detector bandwidth, one can assume $\hat{h}(f) \approx \hat{h}(0) = q$.
Using the relative intensity $\alpha(t)$ from equation~\ref{eq:alphat}, one obtains a simple expression of the power spectral density for a laser detected on a photodiode:
\begin{align}
S^+_I(f) &= 2 S_\epsilon(f) + 2 q^2 N_{avg} +  2 q^2 N_{avg}^2 S_{N/N_{avg}}(f) \nonumber \\
&= S^+_\epsilon(f) + 2 q I_{avg} + I_{avg}^2 S^+_{\alpha}(f)
\label{eq:PSD3}
\end{align}
The relative intensity noise of the photocurrent $I(t)$ reads:
\begin{equation}
\mathcal{RIN}_I(f, I_{avg}) = \frac{S^+_\epsilon(f)}{I_{avg}^2} + \frac{2q}{I_{avg}} + S^+_{\alpha}(f)
\label{eq:RINI}
\end{equation}
It is important to note that $\mathcal{RIN}_I$ covers all current fluctuations while $S^+_{\alpha}(f)$ covers only the classical fluctuations of the laser intensity:
\begin{equation}
\mathcal{RIN}_{\langle I \rangle} = S^+_{\alpha}(f)
\label{eq:Salpha}
\end{equation}
Equation~\ref{eq:PSD3} illustrates the dependence of the electrical PSD at the output of the detector with respect to both frequency and average current from the detector.
The three terms on the right of equations~\ref{eq:PSD3} and~\ref{eq:RINI} are linked to the electrical noise, the laser shot noise, and the laser excess (classical) noise, respectively (Table~\ref{tab:RINlimit}).
The impact of these three terms will be developed further, and measured in the following sections.
\begin{table}
\centering
\begin{tabular}{r | c | c | c }
Limitation \, & \, Electronic \, & \, Shot Noise \, & \, Excess (Classical) Noise \\
\hline
$\mathcal{PSD} \approx$ \, & $S^+_\epsilon(f)$ & $2qI_{avg}$ & $S^+_{\alpha}(f)\, I_{avg}^2$ \\
\hline
$\mathcal{RIN} \approx$ \, & $\frac{S^+_\epsilon(f)}{I_{avg}^2}$ & $\frac{2q}{I_{avg}}$ & $S^+_{\alpha}(f)$ \\
\end{tabular}
\caption{\label{tab:RINlimit}The three measurement regimes and associated dominant power spectral densities and relative intensity noise.}
\end{table}

As can be seen in equation~\ref{eq:SNRRIN}, the SNR for laser intensity measurements is inversely proportional to the RIN integrated over the photodiode bandwidth.
Equation~\ref{eq:RINI} illustrates that the RIN depends on the frequency and average laser intensity.
At low frequencies ($f < 1\,\mathrm{MHz}$) lasers usually features fluctuations in intensity significantly above the shot noise $S^+_{\alpha} \gg 2q/I_{avg}$.
Frequently, this noise renders impossible the direct measurements of small intensity fluctuations ($10^{-4}- 10^{-6}$) induced by SRS.
For this reason, lock-in amplification is typically used to increase the SNR by moving the measurement towards higher frequencies.

\section{Low noise SRS detection through lock-in amplification}
\subsection{System and model description\label{sec:modeldescrip}}
\begin{figure}[htbp]
\includegraphics[width=\textwidth]{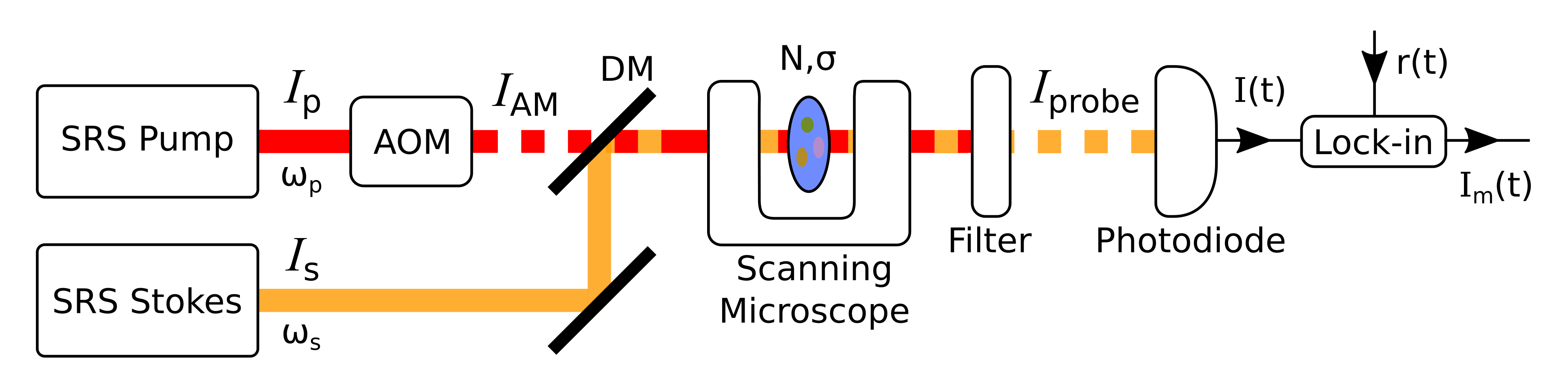}
\caption{\label{fig:SRSsetup} Typical Stimulated Raman Scattering Scanning Microscope system. The SRS pump laser is modulated using an acousto-optic modulator (AOM). It recombines with the Stokes beam using a dichroic mirror (DM) and both are sent on the sample through a scanning microscope. The modulated beam is filtered out using an optical filter and the intensity of the other beam is measured with a photodiode. The signal from the photodiode $I$ is mixed with a reference signal $r$, filtered and amplified using a lock-in amplifier, which generated an output current $I_m$.}
\end{figure}

In a standard SRS system (Figure~\ref{fig:SRSsetup}), a pump laser beam with electric field intensity $\mathcal{I}_{p}$ and a Stokes beam with electric field intensity $\mathcal{I}_{s}$ are sent through a sample.
When collecting $\mathcal{I}_{p}$ or $\mathcal{I}_{s}$ in the far field after their interactions with the sample, one can measure an intensity loss $\Delta \mathcal{I}_{p}$ on the pump beam, and an intensity gain $\Delta \mathcal{I}_{s}$ on the Stokes beam~\cite{Rigneault2018}.
Without loss of generality, the following will assume that the SRS Stokes beam is collected by the photodiode (stimulated Raman gain, SRG, modality), while the SRS Pump beam is discarded using an optical filter.
After interaction with the sample, the Stokes beam has gained a relative intensity $\beta$, proportional to the number of molecular bonds $N$ in the probed volume, their stimulated Raman cross section $\sigma$, and the optical intensity of the SRS pump beam:
\begin{align}
\beta \equiv \frac{\Delta \mathcal{I}_{s}}{\mathcal{I}_{s}} \label{eq:beta}\\
\beta \propto N \sigma \mathcal{I}_{p} \label{eq:betaprop}
\end{align}

The amplitude of the relative SRS gain $\beta$ is typically $10^{-4}$ to $10^{-6}$.
The exact expression of the proportionality factor in equation~\ref{eq:betaprop} depends on many experimental parameters, and in-depth developments can be found in the literature~\cite{Rigneault2018, Cheng2012}.

As a direct result of equations~\ref{eq:beta} and~\ref{eq:betaprop}, if an intensity modulation at frequency $f_0$ is applied on the SRS pump beam (by the AOM), the change in intensity $\Delta \mathcal{I}$ of the Stokes beam will also be modulated at the same frequency.
Here, the first laser (SRS Pump) is referred to as the amplitude modulated (AM) laser, and the second (SRS Stokes) as the demodulated (probe) laser.
Note that the roles (AM/probe) of the two lasers would be switched in the stimulated Raman loss (SRL) modality.
The modulation of the AM beam is assumed to be total:
\begin{equation}
\label{eq:Imod}
\mathcal{I}_{AM} = \mathcal{I}_{p} \frac{1 + \cos(2 \pi f_0 t)}{2}
\end{equation}
The probe laser intensity after interaction on the sample reads:
\begin{equation}
\label{eq:DIprobe}
\mathcal{I}_{probe} = \mathcal{I}_{s} +  \Delta \mathcal{I}_{s} = \mathcal{I}_{s} \Big(1 + \frac{\beta}{2} + \frac{\beta}{2} \cos(2 \pi f_0 t)\Big)
\end{equation}

As described previously, the contribution of SRS to the average optical intensity on the probe beam is not easily detectable.
In order to simplify the mathematical derivations, the DC contribution of the SRS process is discarded, and only on the modulation at frequency $f_0$ is considered.
With this simplification, equation~\ref{eq:DIprobe} becomes:
\begin{equation}
\label{eq:DIprobe2}
\mathcal{I}_{probe} = \mathcal{I}_{s} \Big(1 + \frac{\beta}{2} \cos(2 \pi f_0 t)\Big)
\end{equation}

The probe laser intensity is detected by the photodiode which generates a current, modeled with equation~\ref{eq:photocurrentmodel}.
The difference with the previous derivation (section~\ref{sec:model1}) is that the expected number of electrons generated by the pulse $k$ is now modulated: 
\begin{equation}
N(t_k) =  N_{avg}\alpha(t_k)m(t_k) \label{eq:Ntk}
\end{equation}
Where $\alpha(t)$ is the relative intensity fluctuation in the absence of the modulated beam (equation~\ref{eq:alphat}) and $m(t)$ the modulation transferred from the modulated beam: 
\begin{equation}
m(t) = 1 + \frac{\beta}{2}\cos(2 \pi f_0 t)
\end{equation}
With this definition $I_{avg}m(t)$ fluctuates between $I_{avg} - \frac{\Delta I}{2}$ and $I_{avg} + \frac{\Delta I}{2}$. Where $\Delta I$ is linked to $\beta$ through the following relation:
\begin{equation}
\beta  = \frac{\Delta I}{I_{avg}} 
\end{equation}

The current $I(t)$ flowing from the photodiode is then sent to a lock-in amplifier which mixes it with the reference signal $r(t) = g \, \cos(2 \pi f_0 t)$.
The amplitude $g$ of this reference signal comes as an overall gain and does not affect the discussion on SNR.
The phase of the reference signal is also chosen to be in phase with the modulation signal for optimal demodulation.
The mixed current $I_{m}$ can be expressed as:
\begin{equation}
I_{m}(t) = \Big[ [\sum_k X_k \delta(t - t_k) ] \otimes h(t) + \epsilon(t) \Big] \cdot r(t)
\end{equation}
In terms of Fourier components:
\begin{align}
\hat{I}_{mT}(f) &\mathrel{\mathop{=}\limits_{T\to + \infty}} \hat{I}_{T}(f) \otimes \hat{r}_{T}(f) \label{eq:If} \\
\hat{m}_{T}(f) &= 1 + \frac{\beta}{4} (\delta(f-f_0) + \delta(f+f_0)) \label{eq:mtf} \\
\hat{r}_{T}(f) &= \frac{g}{2} (\delta(f-f_0) + \delta(f+f_0)) \label{eq:rtf}  \\
\hat{m}_{T}(f) \otimes \hat{r}_{T}(f) &\mathrel{\mathop{=}\limits_{T\to + \infty}}  g[\frac{\beta}{4}\delta(f) + \frac{1}{2}(\delta(f-f_0) + \delta(f+f_0)) + \frac{\beta}{8}(\delta(f - 2 f_0) + \delta(f + 2 f_0))] \label{eq:mr}
\end{align}

\subsection{Average current and DC power}
The average current at the output of the lock-in amplifier $I_{m,avg}$ can readily be computed using equation~\ref{eq:Iavg},~\ref{eq:alphat} \ref{eq:If}, and~\ref{eq:mr}:
\begin{align}
I_{m, avg} &= \lim_{T\rightarrow \infty} \frac{1}{T} \langle \hat{I}_{mT}(0) \rangle \nonumber  \\
&= \lim_{T\rightarrow \infty} \frac{1}{T} \langle \hat{I}_{T}(f) \rangle \otimes \hat{r}_{T}(f)\vert_{f=0} \nonumber  \\
&= \lim_{T\rightarrow \infty} \frac{1}{T} I_{avg} \hat{\alpha}_{T}(f) \otimes \hat{m}_{T}(f) \otimes \hat{r}_{T}(f)\vert_{f=0} \nonumber \\
&= \lim_{T\rightarrow \infty} \frac{1}{T} g I_{avg} [(\frac{\beta}{4})\hat{\alpha}_{T}(0) + \frac{1}{2}(\hat{\alpha}_{T}(-f_0) + \hat{\alpha}_{T}(f_0)) + \frac{\beta}{8}(\hat{\alpha}_{T}(- 2 f_0) + \hat{\alpha}_{T}(2 f_0))] \label{eq:Imavg}
\end{align}

The terms $\frac{1}{T}\hat{\alpha}_{T}(\pm f_0)$ and $\frac{1}{T}\hat{\alpha}_{T}(\pm 2 f_0)$ will vanish when averaged over all pulses ($T\rightarrow \infty$).
In the specific case where the modulation frequency $f_0$ is half of the laser repetition rate, the last term will not vanish anymore, and the SNR will effectively be doubled.
Annex~B discusses this scenario, which has been highlighted before by Ozeki and collaborators~\cite{Ozeki2010a}.
In the conventional SRS case, however, the only relevant term is $\lim_{T\rightarrow \infty} \frac{1}{T}\hat{\alpha}_{T}(0) = 1$, and the average mixed current at the output of the lock-in amplifier reads:
\begin{equation}
I_{m, avg} = g I_{avg} \frac{\beta}{4}\label{eq:Imavg2}
\end{equation}

From this current, the DC power at the output of the lock-in with a load resistor $R$ can be derived as:
\begin{equation}
\label{eq:PDC}
P_{DC} = R I_{avg}^2 g^2 \frac{\beta^2}{16}
\end{equation}

\subsection{Power spectral density}
From equation~\ref{eq:If} one can calculate the photocurrent power spectral density $S_{I_m}(f)$ (equation~\ref{eq:SA}), for the mixed current at the output of the lock-in amplifier:
\begin{align}
S_{I_m}(f) &= \lim_{T\rightarrow \infty} \frac{1}{T} \langle \vert\hat{I}_{mT}(f)\vert^2 \rangle \nonumber \\
&= \lim_{T\rightarrow \infty} \frac{1}{T} \langle \vert \hat{I}_{T}(f) \otimes \hat{r}_{T}(f)\vert^2 \rangle \nonumber \\
&= \lim_{T\rightarrow \infty} \frac{1}{T} \langle \vert q \sum_k X_k e^{-2i \pi k f / f_r} \otimes \hat{r}_{T}(f)\vert^2 \rangle  + \lim_{T\rightarrow \infty} \frac{1}{T} \langle \vert\hat{\epsilon}_{T}(f) \otimes \hat{r}_{T}(f)\vert^2 \rangle \label{eq:SIm1}
\end{align}
Detailed calculations can be found in Annex~C. Assuming no correlation between noises at different frequencies, small values of the relative SRS gain ($\beta \ll 1$), and frequencies small compared to the modulation frequency ($f \ll f_0$), the PSD of the lock-in output current can be simplified as:
\begin{equation}
S^+_{I_m}(f) = \frac{g^2}{2} \Big(S^+_{\epsilon}(f_0) + 2 q I_{avg} + I_{avg}^2 S^+_{\alpha}(f_0)\Big) = \frac{g^2}{2} S^+_{I}(f_0) \label{eq:SIm}
\end{equation}

Equation~\ref{eq:SIm} illustrates the advantage of using lock-in detection.
Due to the modulation/demodulation scheme, the low frequency noise features at the output of the lock-in device correspond to the high frequency noise features of the laser.
By choosing the $f_0$ that minimizes these features, one can recover the minimum noise allowed by the laser system.
In particular, by having the modulation frequency $f_0$ sufficiently high, one can avoid the noise at low frequency that is inherent to laser systems.
The factor $\frac{g^2}{2}$ is the gain $G$ of the lock-in amplifier system:
\begin{equation}
G \equiv \frac{g^2}{2}
\label{eq:gain}
\end{equation}

\subsection{Signal to noise ratio}

Integrating the PSD over the lock-in bandwidth $\Delta f$ (equation~\ref{eq:PDf}), the electrical power of the noise (equation~\ref{eq:PNoise}) at the output of the lock-in amplifier can be expressed as:
\begin{equation}
\label{eq:PoutNoise}
P_{Noise} = G \,R \, \Delta f \, S^+_{I}(f_0) = G \,R I_{avg}^2 \,\mathcal{RIN}_I(f_0, I_{avg}) \, \Delta f
\end{equation}

Equation~\ref{eq:PoutNoise} assumes for simplification that the PSD is constant around $f_0$, which is typically a good approximation.
Using equations~\ref{eq:PDC} and~\ref{eq:PoutNoise}, the signal to noise ratio as the output of the lock-in amplifier can ultimately be expressed as:
\begin{equation}
\label{eq:SNR}
SNR = \frac{P_{DC}}{P_{Noise}} = \frac{\beta^2}{8 \Delta f \,\mathcal{RIN}_I(f_0, I_{avg})}
\end{equation}

The result of equation~\ref{eq:SNR} is that, assuming all other sources of noise are negligible, the SNR in a lock-in-based SRS setup is given by the RIN of the laser around the modulation frequency.
Another way to interpret equation~\ref{eq:SNR} is to express the sensitivity of the system, which is the smallest SRS gain $\beta_{min}$ the system can detect with an SNR of 1.
\begin{equation}
\label{eq:betamin}
\beta_{min} = \sqrt{8 \Delta f \,\mathcal{RIN}_I(f_0, I_{avg})}
\end{equation}
As a direct result of equations~\ref{eq:SNR} and~\ref{eq:betamin}, any deviation from the minimal RIN results in a sub-optimal measurement that impairs the SNR, the sensitivity, or slows down signal acquisition.
The time between two successive measurement is inversely proportional to the lock-in bandwidth.
As a result, for a RIN that is doubled, the acquisition time also has to be doubled to maintain the same acquisition SNR.

\subsection{SNR optimization\label{sec:optSNR}}

Combining equations~\ref{eq:SNR} and~\ref{eq:RINI}, the SNR of the system can be written as:

\begin{equation}
\label{eq:SNR2}
\large
SNR(f_0, \Delta f, I_{avg}) = \frac{ \frac{\beta ^ 2}{8 \Delta f}}{\frac{S^+_{\epsilon}(f_0)}{I_{avg}^2} + \frac{2 q }{I_{avg}} + S^+_\alpha(f_0) }
\end{equation}

The different terms of the denominator in equation~\ref{eq:SNR2} are precisely the RIN associated with electronic noise, shot noise, and laser excess noise respectively (Table~\ref{tab:RINlimit}).
Using this equation, one can identify all the parameters that can maximize the SNR in a lock-in-based SRS system:

(i) Maximizing $\beta$, the relative modulation of the probe beam. This goes beyond the scope of this tutorial, but $\beta$ increases linearly with the number of molecular bonds probed, as well as with the laser intensity in the AM beam  (equation~\ref{eq:betaprop}). It is important to note that $\beta$ is constant with respect to the laser intensity in the probed beam, and therefore does not depend on $I_{avg}$. It does, however, depend linearly on the AM laser average intensity, further details are discussed in section~\ref{sec:OptRatio}.

(ii) Minimizing $\Delta f$, the lock-in bandwidth. In principle an arbitrarily high SNR can be achieved by reducing $\Delta f$. In practice the bandwidth has to be large enough to allow changes in the SRS signal. In a SRS point scanning microscope the signal is expected to vary from one pixel to the next. In this case $\Delta f$ should match the pixel acquisition rate to allow fluctuations from one pixel to the next, while filtering out fluctuations within a pixel.

(iii) Maximizing $I_{avg}$ to minimize electrical noise and shot noise. Depending on the photodiode noise features, one can increase the probed laser beam intensity (and therefore $I_{avg}$) to reach the regime where the electronic noise $S^+_{\epsilon}(f_0)/ I_{avg}^2$ becomes negligible with respect to classical and quantum laser fluctuations. In the systems described below, this regime was reached for a few milliwatts of average intensity in the probe beam. When the SNR is limited by the laser, one of two scenarios have to be considered. 

\hspace{0.5cm}a) The RIN is limited by the shot noise, in which case the SNR will increase linearly with $I_{avg}$. The SNR is then limited by how much $I_{avg}$ can be increased. This limit is set by the laser maximum output power, the photodamage threshold, or the laser excess noise.

\hspace{0.5cm}b) The RIN is limited by the laser excess noise, in which case increasing $I_{avg}$ will not change the SNR. The SNR limit allowed by the laser is reached and cannot be improved by increasing the probe laser beam intensity. 

In this last scenario, the laser excess noise ultimately sets the limit on the SNR in a lock-in-based SRS system.
By using a modulation frequency that minimizes this excess RIN one can achieve the optimal SNR.
A proper measurement of the laser RIN is therefore necessary to quantify and optimize the performance of such SRS system.
Two experimental procedures used for laser RIN measurement are detailed in the following section, (i) a complete RIN characterization for different frequencies and optical intensities, and (ii) a faster RIN assessment on an existing lock-in-based SRS microscope.
Both methods are illustrated on two commercial laser sources.

\section{Laser intensity noise measurements}
As described previously the RIN of a photodiode output current is composed of three different terms (equation~\ref{eq:RINI}) that come from electronic noise, quantum fluctuations, and classical fluctuations (Table~\ref{tab:RINlimit}).
The electronic contribution is due only to the photodiode and can be made negligible by choosing the adequate device and laser intensity.
The quantum and classical fluctuations are inherent to the laser source.
Its understanding and characterization is crucial to assess the performance of a SRS system.
The RIN is a function of both frequency and average current provided by the photodiode.
Here, we describe a way to reconstruct the full $\mathcal{RIN}_I(f, I_{avg})$ surface over a large range of frequencies ($500 \, \mathrm{kHz}$ to $30 \, \mathrm{MHz}$) and low laser intensities (0.1mW to 50mW) suitable for SRS imaging applications. 

\begin{figure}[hbtp]
\includegraphics[width=\textwidth]{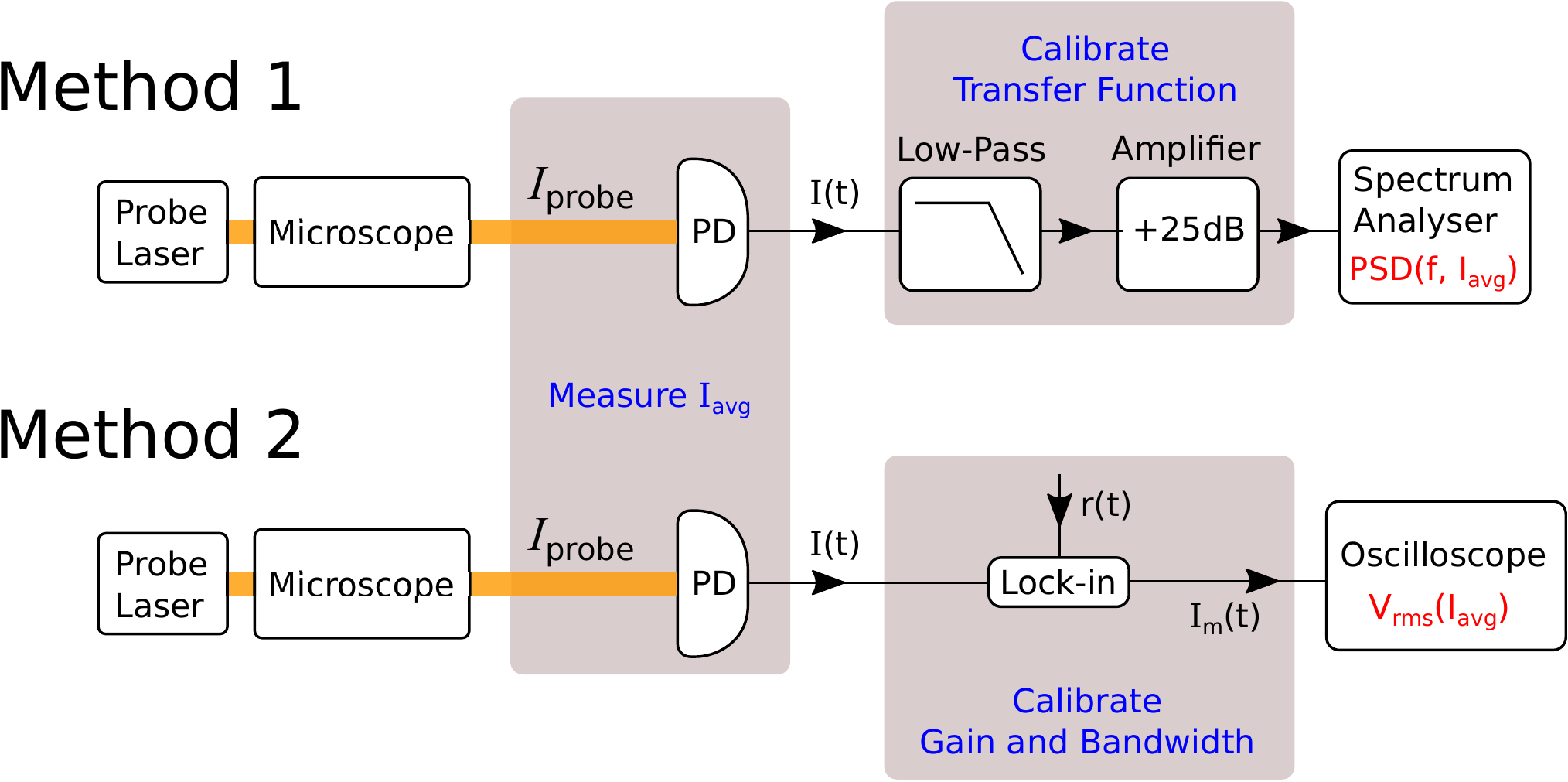}
\caption{\label{fig:methods}Schematics of the RIN measurement methods. In both cases, measurement of the average photocurrent $I_{avg}$ can be performed either directly using an oscilloscope after the photodiode, or estimated by measuring the optical average power using a power meter and using the photoconvertion equation (eq.~\ref{eq:Ioptavg}). Proper characterization of the gain and bandwidth of the measurement electronics is key to properly estimating the laser RIN.}
\end{figure}

\subsection{Measured laser sources\label{sec:sources}}
Device 1 is a Ti:Saph femtosecond laser (Chameleon, Coherent Inc) operating at a repetition rate of \SI{80}{\mega \hertz} with a pulse width duration of \SI{150}{\femto\second} and a wavelength of \SI{800}{\nano \meter}. This beam acts as the (unmodulated) SRS pump beam of a lock-in-based SRS system (SRL modality) whereas the (AM) Stokes beam is provided by a tunable fs optical parametric oscillator OPO (Compact OPO, APE \& Coherent Inc) pumped by the Ti:Saph laser. Here, only the \SI{800}{\nano \meter} beam is considered, as it is the one monitored by the SRS photodiode.

Device 2 is a picosecond OPO (Emerald, APE) operating at a repetition rate of \SI{80}{\mega \hertz} with a pulse duration of \SI{2}{\pico\second} and a wavelength of \SI{800}{\nano \meter}. This OPO is pumped by a 20W frequency doubled ytterbium (Yb) fiber laser (Emerald engine, APE) (\SI{80}{\mega \hertz} rep rate, \SI{2}{\pico\second} pulse duration) that provides also the Stokes beams at \SI{1030}{\nano \meter}. As explained previously, only the \SI{800}{\nano \meter} beam is considered in the RIN measurement. Note that device 2 is similar to the commercially available one box SRS laser system PicoEmerald (APE), although the PicoEmerald uses an 10W frequency doubled Yb fiber laser.  

\subsection{Method 1: frequency and intensity dependence\label{sec:meth1}}
\begin{figure}[hbtp]
\includegraphics[width=0.5 \textwidth]{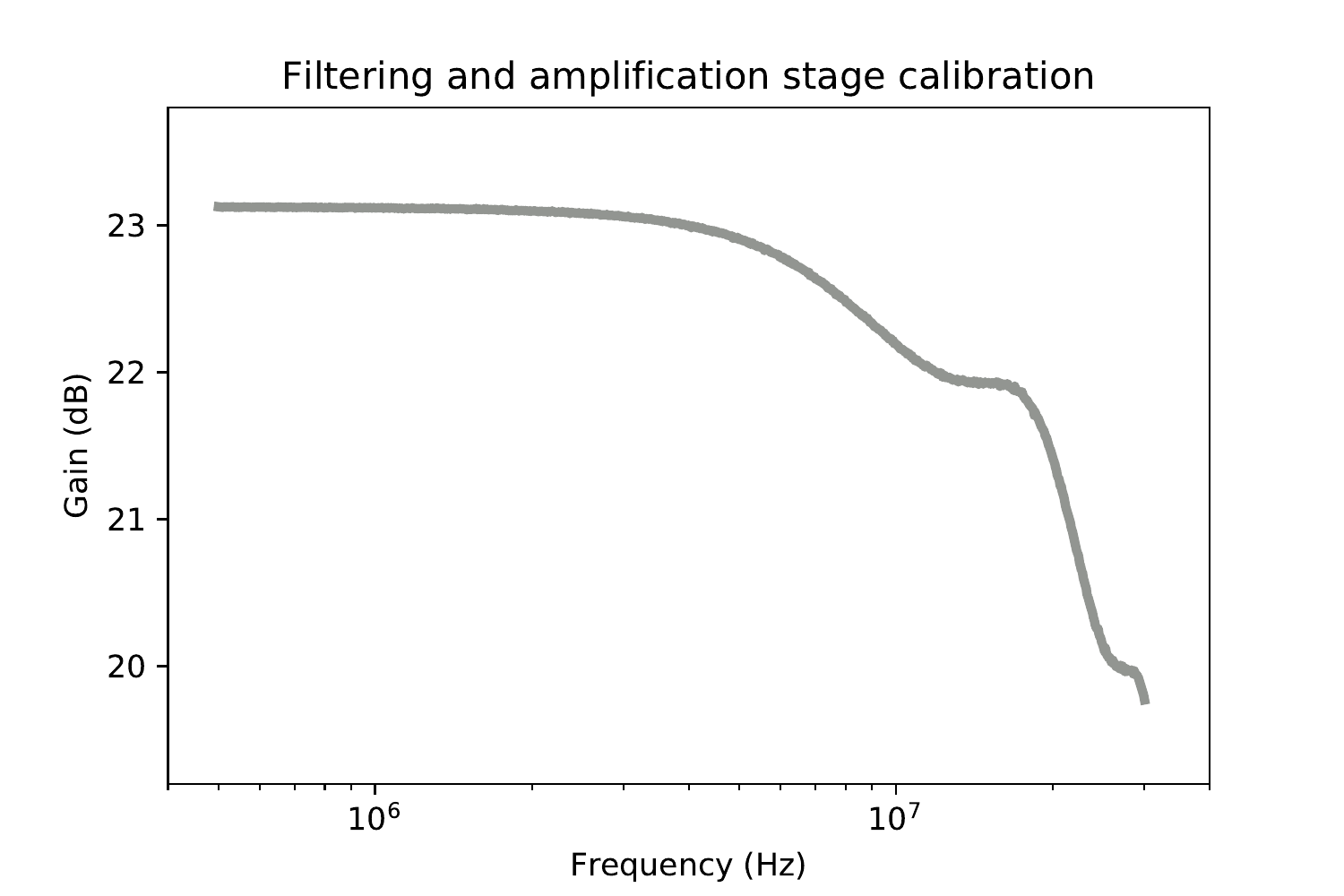}
\caption{\label{fig:filterAmpGain} Supplementary figure: Measured transfer function of the filter and preamplification step.}
\end{figure}

To properly quantify the noise level in devices 1 and 2, fast biased photodiode (Thorlabs DET10A/M, $1 \, \mathrm{ns}$ rise time) is placed at the position of the forward SRS detector (Figure~\ref{fig:SRSsetup}).
The average photocurrent generated by the photodiode was measured with a fast oscilloscope (HP 54111D $500 \, \mathrm{MHz}$ bandwidth, $50\, \mathrm{\Omega}$ input).
The typical photocurrent was several milliamperes.
The associated shot noise PSD can be calculated using Table~(\ref{tab:RINlimit}) and is on the order of $-190 \, \mathrm{dBW/Hz}$ which is far below the input noise ($-180 \, \mathrm{dBW/Hz}$) of the spectrum analyzer used here (Zurich instrument HF2LI).
For this reason, a filtering and amplification step was added.
The photodiode was loaded with a $50\, \mathrm{\Omega}$ resistor and the resulting voltage was filtered with a $30 \, \mathrm{MHz}$ low pass filter (Mini-Circuits BLP$-$30$+$) to damp the 80 MHz repetition rate and harmonics of the lasers.
The signal was then amplified with a \SI{25}{\decibel} low noise preamplifier (Ref 153579, APE) working in the range of $100 \, \mathrm{kHz}$ to $100 \, \mathrm{MHz}$.
The overall effect of this filtering and amplification stage was to amplify by more than $20 \, \mathrm{dB}$ the electrical components between $500 \, \mathrm{kHz}$ and $30 \, \mathrm{MHz}$ (Supplementary Figure~\ref{fig:filterAmpGain}).
Then, the electrical power spectral density was acquired with a spectrum analyzer (Zurich instrument HF2LI).
The PSD was measured for 1000 frequencies from $500 \, \mathrm{kHz}$ to $30 \, \mathrm{MHz}$ with a bandwidth of \SI{1.1}{\kilo \hertz} to \SI{70}{\kilo \hertz}.
The effect of the filter and amplifier was numerically subtracted from the obtained data.

\begin{figure}[hbtp]
\includegraphics[width=0.5 \textwidth]{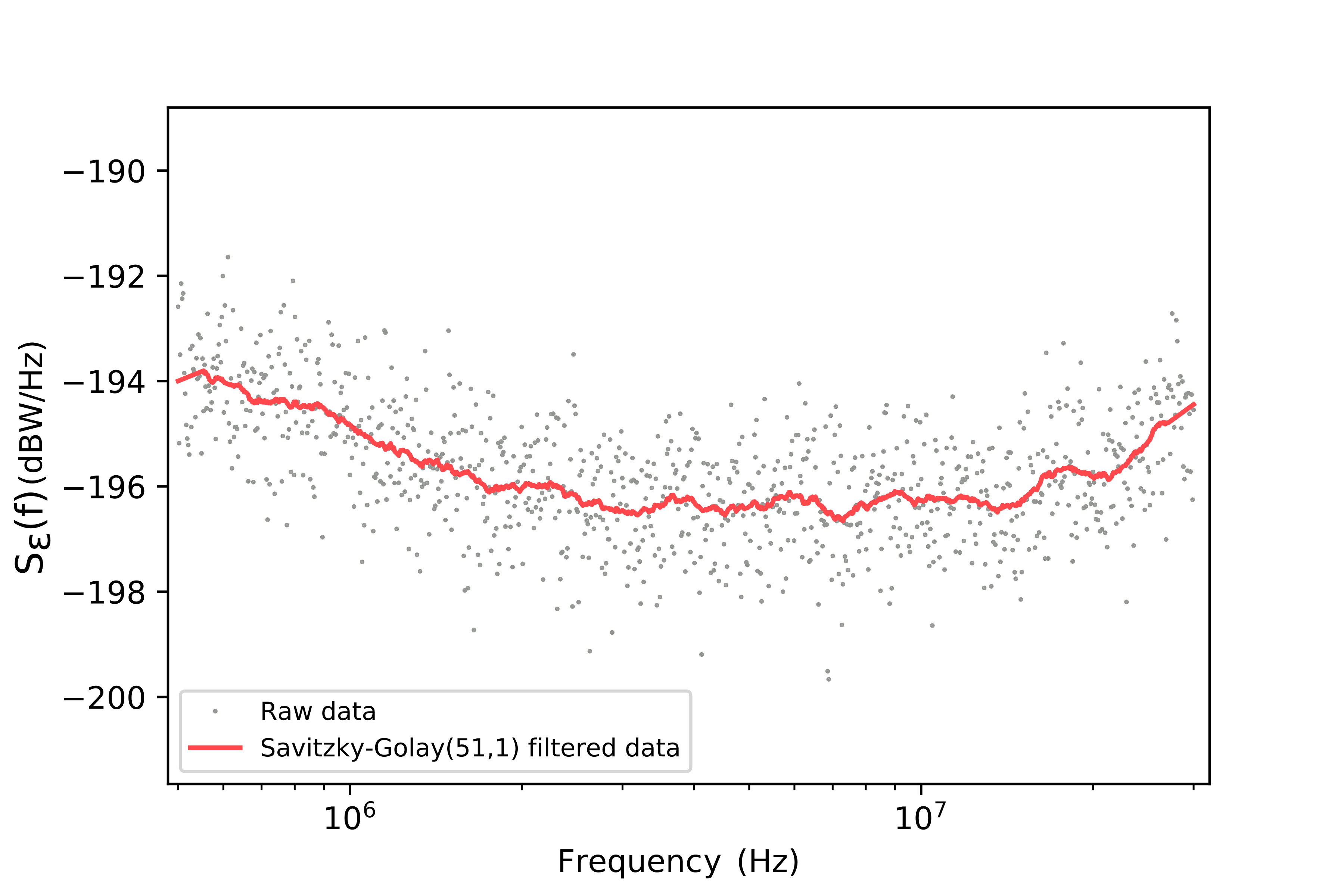}
\caption{\label{fig:elecNoisePSD} Supplementary figure: Equivalent electronic noise power spectral density $S^+_{\epsilon}(f)$.}
\end{figure}

The power spectral density associated with electronic noise was measured by blocking the laser beam. The noise level was $-173 \, \mathrm{dBW/Hz}$ measured at the input of the spectrum analyzer. This noise is a combination of amplified dark current, input and output noise of the amplifier, and input noise of the spectrum analyzer.
All electronic noises are merged in an equivalent power spectral density at the detector $S^+_\epsilon(f)$. The obtained value after compensating for the preamplifier stage gives: $S^+_\epsilon(f) \approx -173 - 23 = -196 \mathrm{dBW/Hz}$, with very little dependence in frequency (Supplementary Figure~\ref{fig:elecNoisePSD}).
After numerically accounting for filtering and amplification, the contribution of the electronic noises to the power spectral density was removed numerically to keep only the laser RIN:
\begin{equation}
\mathcal{RIN}_{laser}(f, I_{avg}) = \mathcal{RIN}_I(f, I_{avg}) - \frac{S^+_\epsilon(f)}{I_{avg}^2} = \frac{2q}{I_{avg}} + S^+_{\alpha}(f)
\label{eq:RINlaser}
\end{equation}

\subsection{Method 1: Results}
\begin{figure}[hbtp]
\includegraphics[width=\textwidth]{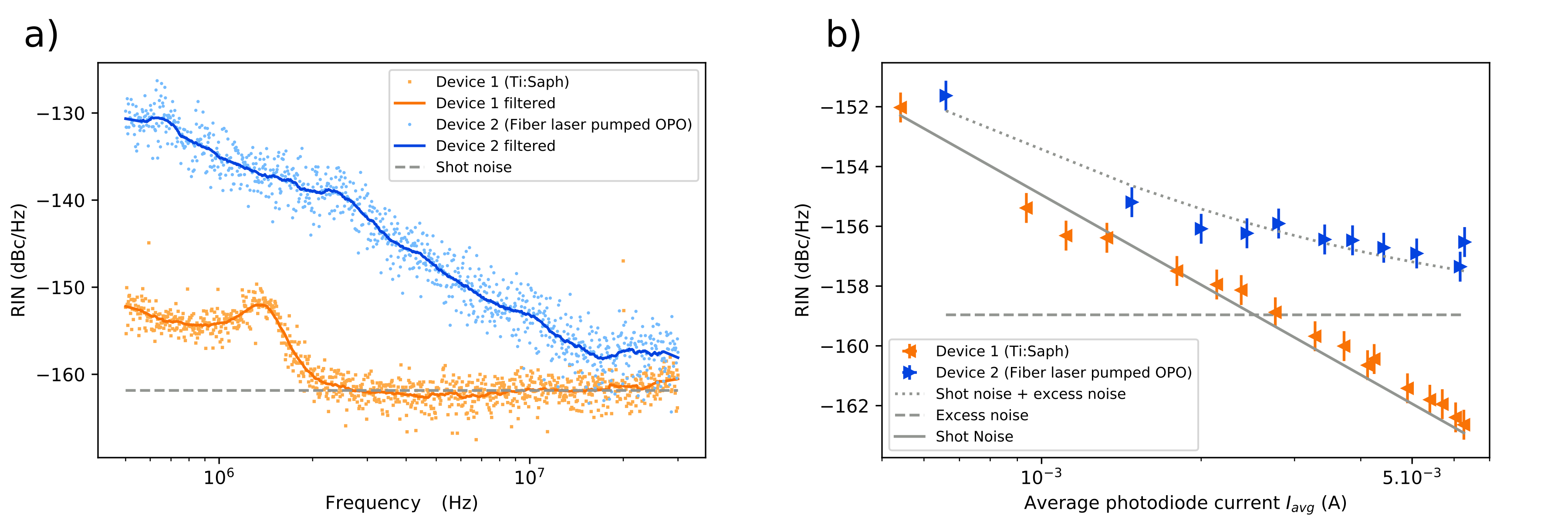}
\caption{\label{fig:RINvsFreqvsCurrent} a) Relative intensity noise measured as a function of frequency for device 1 (Ti:Saph) and 2 (Fiber laser pumped OPO) with average photodiode current $I_{avg} = 5 \, \mathrm{mA}$. The solid line are the filtered data (Savitzky–Golay filter, 51 points, 1st order). b) Relative intensity noise (RIN) as a function of the average photodiode current for device 1 (Ti:Saph, left triangles) and 2 (Fiber laser pumped OPO, right triangles). The RIN was averaged between 10 and $30 \, \mathrm{MHz}$. The excess noise on device 2 (dashed line) is fitted to match the measurements. The electronic noise was removed numerically.}
\end{figure}

For an average photocurrent of $I_{avg} = 5 \, \mathrm{mA}$ the recovered laser RIN are plotted in Figure~\ref{fig:RINvsFreqvsCurrent}a for devices 1 and 2.
For device 1 (Ti:Saph laser), the measured data matches the expected shot noise limit for frequencies above $2 \, \mathrm{MHz}$.
Consequently, any SRS related modulation of the laser above this frequency can be detected down to the shot noise limit.
For device 2, the measured data is within $5 \, \mathrm{dB}$ of the shot noise limit for frequencies above $20 \, \mathrm{MHz}$.
This corresponds to a SNR for the system - at such modulation frequencies and photocurrent - that is only a factor of 3 below the shot noise limit, meaning that a measurement requires a bandwidth 3 times smaller than that of a shot noise limited system to receive the same SNR (equation~\ref{eq:SNRRIN}). Note that for device 2, pumping the OPO with a 10W Yb fiber laser (PicoEmerald case) would result in a $3 \, \mathrm{dB}$ lower RIN, bringing the OPO 2dB above the shot noise for frequencies above $20 \, \mathrm{MHz}$.

\begin{figure}[hbtp]
\includegraphics[width=\textwidth]{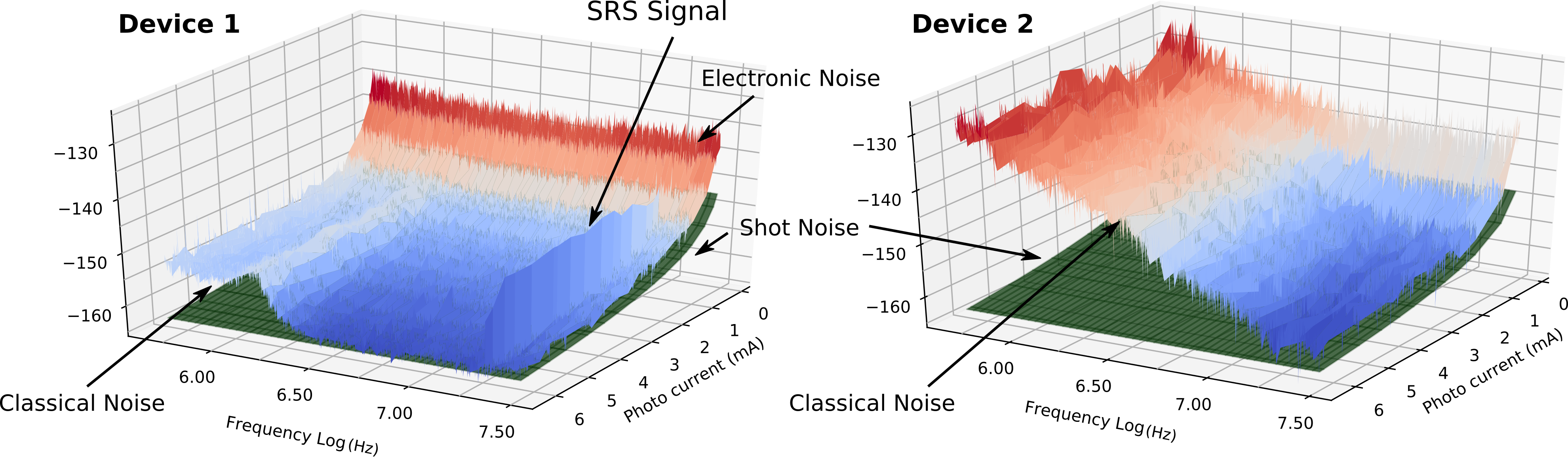}
\caption{\label{fig:RINSurfaces} RIN of device 1 (left) and 2 (right) as a function of frequency ranging from $500 \, \mathrm{kHz}$ to $30 \, \mathrm{MHz}$ and photocurrents ranging from $0.5 \, \mathrm{mA}$ to $7 \, \mathrm{mA}$. The noise floor set by the shot noise is represented by the green bottom surface. At low photocurrents the electronic contribution to the RIN diverges, at low frequencies the classical noise dominates. The peak shown on device 1 at 20MHz corresponds to a SRS induced modulation at this frequency.}
\end{figure}

Complementary to the frequency analysis, one needs to measure the RIN for different photodiode currents to characterize the optimal setup working parameters.
This analysis was performed by recording the RIN as a function of frequency for photocurrents ranging from $0.5 \, \mathrm{mA}$ to $7 \, \mathrm{mA}$, after which the detector saturates.
This saturation comes in the form of a distorted signal from the photodiode output current visualized on the oscilloscope.
The surfaces $\mathcal{RIN}(f, I_{avg})$ are plotted in Figures~\ref{fig:RINSurfaces} for both devices.
To better illustrate $\mathcal{RIN}(f=20 \mathrm{MHz}, I_{avg}$), its value was averaged from 10 to $30 \, \mathrm{MHz}$ in the frequency domain (Figure~\ref{fig:RINvsFreqvsCurrent}b).

Device 1 is shot noise limited around $20 \, \mathrm{MHz}$ for all photodiode current, and therefore all laser intensities, allowed by the photodiode.
Device 2 shows a small excess noise of $5 \, \mathrm{dB}$ above the shot noise for $5 \, \mathrm{mA}$ photocurrent.
As expected from the model, device 2 is shot noise limited for small enough photocurrents ($<$\SI{500}{\micro \ampere}) and reaches a plateau at $S_\alpha(f_0 = 20\, \mathrm{MHz}) = -159 \, \mathrm{dBc/Hz}$.
Supplementary information provides a practical handbook and software matlab script to implement Method 1.

\subsection{Method 2: using an existing lock-in SRS system}
Although the measurement procedure described above provides a complete measurement of the laser RIN, it requires low noise filtering and amplification steps, as well as a sensitive spectrum analyzer. 
When characterizing the performance of a lock-in-based SRS system, such measurement can be simplified by making use of the existing lock-in amplifier.
As can be seen from equation~\ref{eq:PNoise} and~\ref{eq:PoutNoise}, the electrical power $P_{Noise} = \mathrm{Var}[V_{m}] / R$ of the voltage fluctuations at the output of a lock-in amplifier is a direct measurement of the photodiode current PSD (and therefore RIN) around the reference frequency of the lock-in:
\begin{align}
S^+_{I}(f_0) &= \frac{P_{Noise}}{G \,R \, \Delta f} = \frac{\mathrm{Var}[V_{m}]}{G \, R^2 \, \Delta f} \label{eq:SImeth2} \\
\mathcal{RIN}_I(f_0, I_{avg}) &= \frac{\mathrm{Var}[V_{m}]}{G \, \Delta f \, R^2 \,I_{avg}^2}  = \frac{S^+_\epsilon(f_0)}{I_{avg}^2} + \frac{2q}{I_{avg}} + S^+_{\alpha}(f_0)\label{eq:RINmeth2}
\end{align} 

The voltage variance can be measured directly using an oscilloscope with a bandwidth larger than that of the lock-in.
Proper characterization of the lock-in gain $G$ and bandwidth $\Delta f$ is important to measure the parameters in equation~\ref{eq:RINmeth2} and ensure a quantitative comparison of the model with experimental data.
In this study, a commercially available detector and lock-in amplifier (APE LIA, Berlin) was used.
The integration $\tau$ time was set to \SI{20}{\micro \second} which is linked to its bandwidth $\Delta f$ by:
\begin{equation}
\label{eq:Bandwidth}
\Delta f = \frac{1}{2 \pi \tau} = 8 \, \mathrm{kHz}
\end{equation}

The gain $G$ of the lock-in amplifier was measured experimentally using a reference signal at $f_0 = 20 \, \mathrm{MHz}$. The input signal entering the lock-in was a sinusoidal voltage of amplitude $300 \, \mathrm{uV}$ and frequency $f_0$. The output signal from the lock-in resulted in a DC voltage of amplitude $1.8 \, \mathrm{V}$. The associated gain in power was therefore measure to be $G = 20 \log ( V_{out} / V_{in} ) =  75.56 \, \mathrm{dB}$. The reference, lock-in input and output signals were generated and recorded using a multipurpose system (Zurich Instrument HF2LI).

Given $G$ and $\Delta f$, and following equation~\ref{eq:RINmeth2}, the RIN of the photodiode current $I$ can be measured experimentally using the lock-in, by recording the variance of the lock-in output voltage.
This RIN measurement is made at the lock-in frequency $f_0$ and varying laser average intensity (and therefore $I_{avg}$).

More specifically, by plotting the RIN as a function of $I_{avg}$ on a log-log plot, one can identify the different regimes (Electronic noise, Shot noise, Classical noise) of limitation of the measurement.

The electronic noise $S^+_\epsilon(f_0)$ can be measured directly by blocking the laser beam before the photodiode ($I_{avg} = 0$).
The measured lock-in output voltage standard deviation in the absence of the laser was $\sqrt{\mathrm{Var}[V_{m}]} = 1.1 \, \mathrm{mV}$. This corresponds to an electrical noise power in a $R=50\, \mathrm{Ohm}$ resistance of:
\begin{equation}
P_{Noise} = \mathrm{Var}[V_{m}] / R = 2.42 \times 10^{-8} \, \mathrm{W} = -76 \, \mathrm{dBW}
\end{equation} 
Following equation~\ref{eq:SImeth2}, the electrical noise prior to amplification is therefore:
\begin{equation}
S^+_\epsilon(f_0) = -190 \, \mathrm{dBW/Hz} 
\end{equation} 

The PSD and RIN associated to laser shot noise are given by Table~\ref{tab:RINlimit}.
The average photocurrent was measured through the average optical intensity using equation~\ref{eq:Ioptavg}, knowing the quantum efficiency $\eta = 0.8$ of the photodiode at \SI{800}{\nano \meter}. As a result for a laser at $800\, \mathrm{nm}$ wavelength $I_{avg} = 0.5 \_ \mathcal{I}_{opt, avg}$ where $\mathcal{I}_{opt, avg}$ is expressed in Watts and $I_{avg}$ in Amperes.
The RIN was measured on the two commercial laser systems described in Section~\ref{sec:sources}, using the lock-in voltage output standard deviation, for laser average optical intensities ranging from $0.6$ to $50 \, \mathrm{mW}$ on the photodiode (Figure~\ref{fig:RINVSoptPower}).
The contribution of the electronic noise was removed numerically, as described in section~\ref{sec:meth1}.

\subsection{Method 2: Results\label{sec:RINResults}}
The measured RIN from device 1 matches the model (dotted orange line), with no measurable classical (excess) laser noise. For photocurrents $I_{avg}$ higher than $5 \, \mathrm{mA}$ (i.e. laser intensities on the photodiode above $10 \, \mathrm{mW}$) the electronic noise is negligible and the SRS measurement is shot noise limited.

\begin{figure}[htbp]
\includegraphics[width=\textwidth]{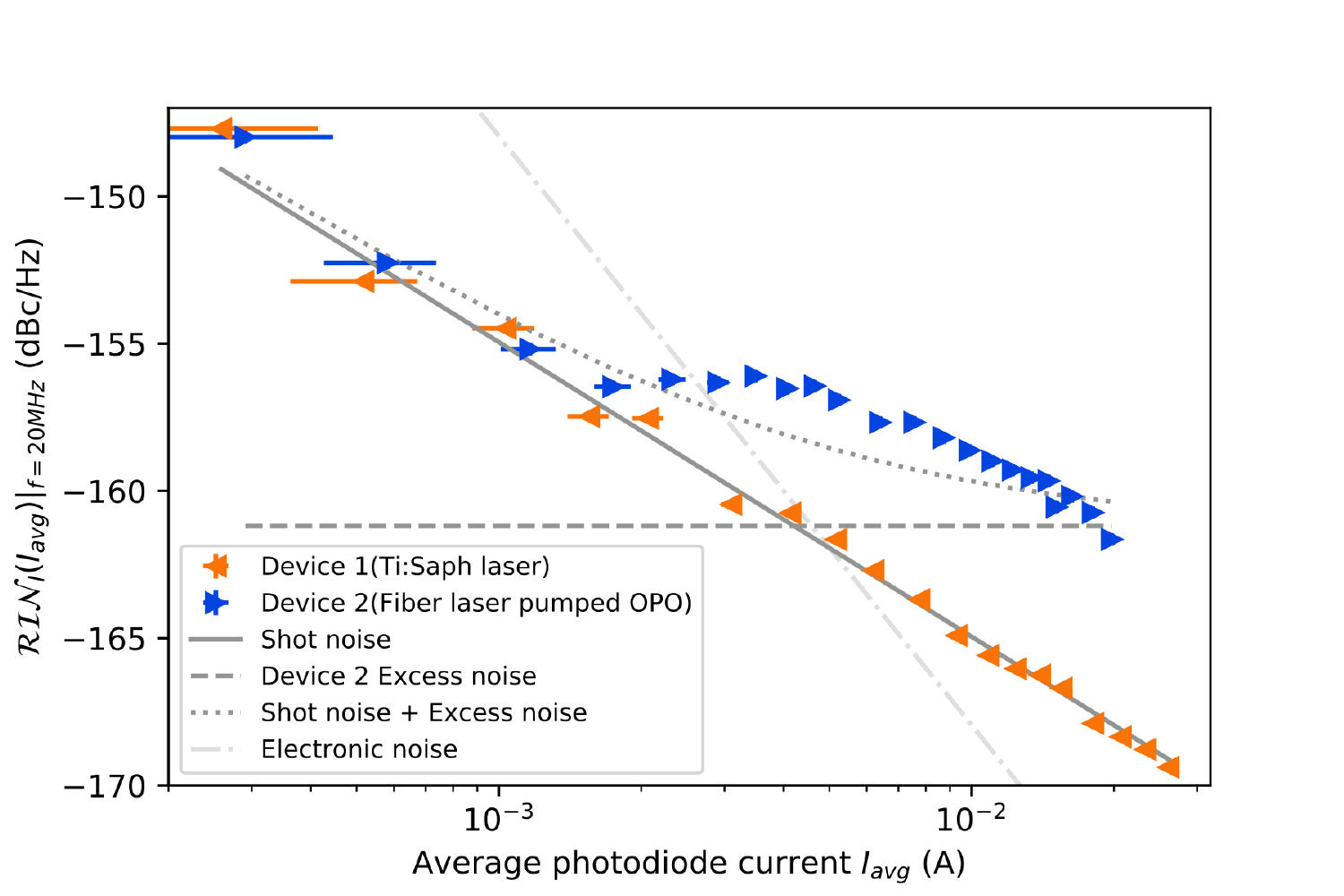}
\caption{\label{fig:RINVSoptPower} Photodiode current RIN (expressed in dBW), for fixed frequency $f=20 \, \mathrm{mW}$ and increasing average laser photodiode current $I_{avg}$ (and therefore average laser intensity). The figure shows measurements for device 1 (Ti:Saph - orange left triangles) and device 2 (Fiber laser pumped OPO - blue right triangles). The shot noise (solid line) and excess laser noise (dash line) are added to provide the modeled noise for device 2 (dotted line). The electronic noise was removed numerically from all measurements but is shown (dash-dotted line) for comparison.}
\end{figure}

Excess laser noise ($S_\alpha$) has to be introduced to fit the experimental data from device 2 with the model (equation~\ref{eq:RINmeth2}).
This noise is estimated using the RIN measurements for high $I_{avg}$ values:
\begin{equation}
S_{\alpha, dev2}(f=20 \, \mathrm{MHz}) = -161 \, \mathrm{dBc/Hz}
\end{equation}
Using this value of excess laser RIN, the measured RIN from device 2 can be fitted with the model (Figure~\ref{fig:RINVSoptPower}).

The RIN measurement obtained using Method 2 (existing lock-in amplifier) is slightly discordant with the more complete study using Method 1.
The excess laser RIN obtained is $2 \, \mathrm{dB}$ below the previously measured value.
This difference can be attributed to changes in the OPO alignment that introduced additional excess laser noise in the first case.
It is noted that Method 2 measurement of the RIN, for device 2, presents relatively strong discrepancies with the model, in the form of fluctuations above and below the expected value.
These fluctuations are also attributed to changes in the OPO alignment during the time of the measurement, which introduced up to $2 \, \mathrm{dB}$ of fluctuations in the RIN measurement.

\section{Discussion}
\subsection{Consequences of RIN on SRS measurements}
Excess laser RIN has a direct and major impact on the sensitivity of lock-in-based SRS measurements.
As illustrated by equation~\ref{eq:SNR2}, the SNR in such measurement is limited by either electronic noise, shot noise, or laser excess intensity noise.
While electronic noise can usually be avoided, and shot noise sets the physical limit, excess laser RIN is more difficult to control and minimize.
Such excess noise in lasers can usually be explained by excess noise in the pump laser or amplified stimulated emission, which are harder to address.

Since the RIN is a function of frequency, the choice of the laser modulation frequency used in the lock-in detection is an important parameter to minimize the laser RIN and maximize the SNR.
As illustrated in Figure~\ref{fig:RINvsFreqvsCurrent}a, the RIN can be lowered by an order of magnitude by simply changing the modulation frequency from 5MHz to 20MHz, therefore increasing the SNR by a factor of 10.

It is crucial to note that the SNR is limited by the laser that is used as a probe and detected by the photodiode.
In the case of SRS this is particularly important, since only one of the laser used (pump or stokes) is the probe and has to exhibit a low noise.
The other laser is usually amplitude modulated and its contribution to noise is negligible.
Complementary to this, once the SNR is limited by excess laser RIN, there is no advantage of increasing the laser power at the sample plane (and therefore the current of the photodiode).
Since the SRS signal scales as the laser noise in this case, the SNR will not be improved by using higher laser power, as detailed in section~\ref{sec:optSNR}.
Although it is important to understand how the SNR can be improved by increasing the intensity of the SRS pump or Stokes lasers, this analysis should be done considering the photodamage on the sample, that usually sets the experimental limit to the laser intensities. 

\subsection{Optimal pump probe ratio\label{sec:OptRatio}}
It has been demonstrated that if the SNR of a lock-in-based system is shot noise limited, and if the photodamage is linear with the intensity of both laser beams, the optimal SNR is achieved for constant photodamage when the AM beam has twice the intensity of the probe beam\cite{Moester2015}.

This result can be extracted from equation~\ref{eq:SNR2}. The total optical intensity on the sample is:
\begin{equation}
    \mathcal{I}_{tot} = \mathcal{I}_{probe} + \mathcal{I}_{AM}
\end{equation}
and the SNR of the measurement is, for a certain proportionality constant $\kappa$:
\begin{equation}
    \mathcal{SNR}(\mathcal{I}_{AM}, \mathcal{I}_{probe}) = \kappa \mathcal{I}_{AM}^2 (\mathcal{RIN}_{I}(\mathcal{I}_{probe}))^{-1} \label{eq:SNR4}
\end{equation}
The optimal ratio between the probe and AM laser intensities is obtained by deriving equation~\ref{eq:SNR4} (or it's logarithm) with respect to $\mathcal{I}_{AM}$, while keeping $\mathcal{I}_{tot}$ constant.
Omitting electronic noises, the resulting optimal relationship between the intensities becomes:
\begin{equation}
   \mathcal{I}_{AM} = 2 \, \mathcal{I}_{probe} + S^+_{\alpha}(f_0) \frac{\eta}{h \nu}\mathcal{I}_{probe}^2 \label{eq:OptRatio}
\end{equation}
Note that this relationship is expressed in terms of average intensity, and that the peak intensity of the AM beam is twice as high due to modulation.
The result is consistent with the 2:1 (AM/probe) ratio in the case of shot noise limited measurements ($S^+_{\alpha}(f_0) \ll 2q/ I _{avg}$), and an even more biased ratio towards the AM beam in the case of a noisy probe beam ($S^+_{\alpha}(f_0) \geq 2q/ I _{avg}$), as discussed above.
Equation~\ref{eq:OptRatio} generalizes the result obtained by Moester and collaborators\cite{Moester2015}, in the case of excess laser intensity noise.

\subsection{Sensitivity and performance report}
In the work presented here, using device 2 as the probe beam, there is no need to push the intensity above few milliwatts (on the photodiode).
In this case it is interesting to increase the optical intensity on the second (modulated) SRS laser to increase the SNR.

With a lock-in integration time of $\tau = 20 \, \mathrm{\mu s}$ and average photocurrent of $I_{avg} = 5 \, \mathrm{mA}$ (i.e. $10 \, \mathrm{mW}$ of laser power on the photodiode) the measured RIN (Figure~\ref{fig:RINvsFreqvsCurrent}) is -162 and -157 $\mathrm{dBc/Hz}$ for devices 1 and 2, respectively. The SRS detection limit for device 1 is therefore predicted to be (equation~\ref{eq:betamin}):
\begin{equation}
\label{eq:DeltaIoverIdev1}
\frac{\Delta I_{min}}{I_{avg}} = \beta_{min} = \sqrt{\frac{8}{2 \pi \tau} 10^{-16.2}} = 2.0 \times 10^{-6}
\end{equation}

And for device 2:
\begin{equation}
\label{eq:DeltaIoverIdev2}
\frac{\Delta I_{min}}{I_{avg}} = \beta_{min} = \sqrt{\frac{8}{2 \pi \tau} 10^{-15.7}} = 3.6 \times 10^{-6}
\end{equation}

To reach such sensitivity with a fiber laser usually requires noise cancellation using balanced detection schemes~\cite{Crisafi2017, Nose2012a, Freudiger2014a}.
Because the sensitivity of the system also depends on its bandwidth, the proper way to evaluate and assess the performance of a SRS system with respect to laser noise is to specify:

(i) The average photocurrent from the detector, which sets the shot noise limited RIN and SNR (Table~\ref{tab:RINlimit}).

(ii) The achieved experimental RIN by either its absolute value or its distance in dB from the shot noise limit.

\subsection{Future developments}
When imaging biological samples and other diluted species, the number of Raman active molecules in the focal spot is such that to detect a signal, one needs to stay on a pixel from several to tens of microseconds.
The associated bandwidth is therefore limited to tens of kilohertz, and the fastest one can expect to perform imaging is tens of thousands of pixels per second.
This acquisition rate corresponds to a few frames per second with 100 by 100 pixel images.
This limit could be overcome in two ways.

First, by reducing the laser repetition frequency and increasing the pulse peak power.
This would further increase the efficiency of the SRS process and render modulation at half of the repetition frequency more practical.
In this case low frequency noise would be extremely important to characterize in order to preserve the SNR.
The limitation set by nonlinear photodamage will however put a limit of the amount of peak power that can be delivered on the sample.

Second and complementary to this first point, as the time spent per pixel cannot be diminished further, major developments are to be expected in spatially multiplexed SRS.
Either by having multiple foci at once in the sample\cite{Heuke2018b}, or with more robust spatial multiplexing, one can increase the amount of information collected from the sample and further increase the imaging speed.
In this case, a lot of power will be necessary to enable the SRS process to take place at multiple locations.
This high power can be achieved with fiber lasers, or optical parametric amplifiers, but the question of noise in such system still needs to be addressed to ensure optimal - shot noise limited - SNR.

\section{Conclusion}
This work presented a description and ways to characterize laser noise in the context of stimulated Raman scattering. 
The laser excess RIN ($S_\alpha(f)$) was introduced as well as the shot noise RIN.
The relation between RIN and signal to noise ratio in SRS was derived (equation~\ref{eq:SNR}), as well as the optimal ratio of pump and Stokes intensities to maximize the SNR for non shot noise limited systems.
Two methods for measuring the RIN were presented, either with a full characterization using commercially available electronic parts, or a more accessible implementation using the SRS lockin amplifier.
Two laser systems were studied.
One was a solid state Ti-Sapphire laser which was shot noise limited around $20 \, \mathrm{MHz}$ for laser power up to $70 \, \mathrm{mW}$.
The second system was an OPO pumped by a 20W fiber laser which showed excess RIN of $-160 \pm 1 \, \mathrm{dBc/Hz}$ at $20 \, \mathrm{MHz}$.
Future developments in SRS will likely involve an improvement in the noise performance of lasers, and a systematic characterization of the system's noise.

\begin{acknowledgments}
We would like to acknowledge Professor Randy A. Bartels for his useful input on this work.
Dr. Ingo Rimke has financial interest in APE.
\end{acknowledgments}

\section{Annex A}
A more detailed derivation of the photocurrent power spectral density for the light detection model presented in this work has been done by Quinlan and collaborators~\cite{Quinlan2013}.
The concise calculation using the notations used here gives:
\begin{align}
S_I(f) &= \lim_{T\rightarrow \infty} \frac{1}{T} \langle \vert\hat{I}_T(f)\vert^2 \rangle  \nonumber \\
&= \lim_{T\rightarrow \infty} \frac{1}{T} \langle \vert \hat{h}_{T}(f) [\sum_k X_k e^{-2i \pi k f / f_r} ] + \hat{\epsilon}_{T}(f)\vert^2 \rangle \nonumber \\
&= S_\epsilon(f) + \lim_{T\rightarrow \infty} \frac{\vert \hat{h}_{T}(f)\vert^2}{T}  \sum_k \sum_l \langle X_k X_k \rangle e^{-2i \pi (k-l) f / f_r} \nonumber \\
&= S_\epsilon(f) + \lim_{T\rightarrow \infty} \frac{\vert \hat{h}_{T}(f)\vert^2}{T} [ \sum_k \sum_l N(t_k) N(t_l) e^{-2i \pi (k-l) f / f_r} + \sum_k N(t_k) ]\nonumber \\
&= S_\epsilon(f) + \vert \hat{h}(f)\vert^2 [ S_N(f) + N_{avg} ]\nonumber
\end{align}

\section{Annex B}
As explained in more details in the work of Ozeki and collaborators~\cite{Ozeki2010a}, using a modulation frequency that is half of the repetition rate of the laser can increase by two fold the SNR.
In this particular case, we obtain:
\begin{align}
\frac{1}{T} \hat{\alpha}_{T}(\pm 2 f_0) &= \frac{1}{T} \int^{T/2}_{-T/2} \frac{\langle I(t) \rangle}{I_{avg}}e^{\pm 4 i \pi f_0 t}\mathrm{d}t \nonumber \\
&= \frac{1}{T} \int^{T/2}_{-T/2}  \sum_k \frac{N(t_k)}{q N_{avg}} h(t - t_k)  e^{\pm 4 i \pi f_0 t}\mathrm{d}t \nonumber
\end{align}
With the approximating that $h(t)$ is a Dirac distribution of area q:
\begin{align}
&\mathrel{\mathop{=}\limits_{T\to + \infty}} \frac{1}{T} \sum_{t_k \in [-T/2,T/2]} \frac{N(t_k)}{N_{avg}} e^{\pm 2 i k \pi (\frac{2f_0}{f_r})} \nonumber 
\end{align}

The sum does not vanish when $2 f_0 = f_r$, instead the terms add up constructively and the limit give:
\begin{equation}
\lim_{T\rightarrow \infty} \frac{1}{T} \hat{\alpha}_{T}(\pm 2 f_0) = \lim_{T\rightarrow \infty} \frac{1}{T} \hat{\alpha}_{T}(0) = 1
\end{equation}
Adding the contribution of this terms to equation~\ref{eq:Imavg}, the mixed average current $I_{m, avg}$ at the output of the lockin amplifier is doubled, and the DC power is quadrupled.
The terms $\hat{\alpha}_{T}(\pm 2 f_0)$ are also involved in the calculation of the noise Power (Annex~C, equation~\ref{eq:SImannex}).
Taking those into account, the noise power is doubled as well, resulting in a net improvement of the SNR of a factor of 2.

\section{Annex C}
Equation~\ref{eq:SIm1} is valid when the electronic noise has a zero expected value and is independent from the laser intensity noise.
The second term of equation~\ref{eq:SIm1} correspond to electronic noise and can be expressed as:
\begin{equation}
\lim_{T\rightarrow \infty} \frac{1}{T} \langle \vert\hat{\epsilon}_{T}(f) \otimes \hat{r}_{T}(f)\vert^2 \rangle = \frac{g^2}{4}[S_{\epsilon}(f-f_0) + S_{\epsilon}(f+f_0)] \label{eq:Se1}
\end{equation}
The first term of equation~\ref{eq:SIm1} relates to laser noise and requires more computation:

\begin{align}
&\langle \vert q \sum_k X_k e^{-2i \pi k f / f_r} \otimes \hat{r}_{T}(f)\vert^2 \rangle \nonumber \\
= \, &\langle \mathcal{F}_T \{ q \sum_k X_k \delta(t-t_k) r(t) \}(f) \mathcal{F}_T^\star \{ q \sum_l X_l \delta(t-t_l) r(t) \}(f)\rangle \nonumber \\
= \, & q^2 \sum_k \sum_l \langle X_k X_l \rangle \mathcal{F}_T \{ \delta(t-t_k) r(t) \}(f) \mathcal{F}_T^\star \{ \delta(t-t_l) r(t) \}(f) \nonumber \\
\end{align}
Using equations~\ref{eq:PkPkk} and~\ref{eq:Ntk}:
\begin{align}
= \, & q^2 \sum_k \sum_l \Big( N_{avg}^2 \alpha(t_k) m(t_k) \alpha(t_l) m(t_l) + N_{avg} \alpha(t_k) m(t_k) \delta_{k,l} \Big) r(t_k) e^{-2 i \pi k f / f_r } r(t_l) e^{2 i \pi l f / f_r }  \nonumber \\
= \, & I_{avg}^2 \vert \sum_k \alpha(t_k) m(t_k) r(t_k) e^{-2 i \pi k f / f_r } \vert^2 + q I_{avg} \sum_k \alpha(t_k) m(t_k)  \vert r(t_k) \vert ^2 \nonumber \\
\end{align}
Using the property that the Fourier Transform of a product is the convolution of the Fourier Transforms:
\begin{align}
= \, & I_{avg}^2 \vert \hat{\alpha}_{T}(f) \otimes \hat{m}_{T}(f) \otimes \hat{r}_{T}(f)\vert^2 + q I_{avg} \hat{\alpha}_{T}(f) \otimes \hat{m}_{T}(f) \otimes \hat{r}_{T}(f) \otimes \hat{r}_{T}(f) \vert_{f=0} \nonumber \\
\end{align}
Using the definition of $\hat{m}_{T}(f)$ and $\hat{r}_{T}(f)$ (equations~\ref{eq:mtf} and~\ref{eq:rtf}):
\begin{align}
= \, & g^2I_{avg}^2 \vert \hat{\alpha}_{T}(f) \otimes \Big( \frac{\beta}{4}\delta(f) + \frac{1}{2}(\delta(f-f_0) + \delta(f+f_0)) + \frac{\beta}{8}(\delta(f - 2 f_0) + \delta(f + 2 f_0) \Big) \vert^2 \nonumber \\
& \hspace{1cm} + q g^2 I_{avg} \hat{\alpha}_{T}(f) \otimes \Big( \frac{\beta}{8}(\delta(f-f_0) + \delta(f+f_0)) + \frac{1}{2}\delta(f) + \frac{1}{4}(\delta(f-2f_0) + \delta(f+2f_0)) \nonumber \\
& \hspace{1cm} + \frac{\beta}{16}(\delta(f - f_0) + \delta(f + f_0))+ \frac{\beta}{16}(\delta(f - 3 f_0) + \delta(f + 3 f_0)) \Big) \vert_{f=0} \nonumber \\
= \, & \frac{g^2I_{avg}^2}{4} \vert \frac{\beta}{2}\hat{\alpha}_{T}(f) + \hat{\alpha}_{T}(f+f_0) + \hat{\alpha}_{T}(f-f_0) + \frac{\beta}{4} (\hat{\alpha}_{T}(f+2f_0) + \hat{\alpha}_{T}(f-2f_0)) \vert^2 \nonumber \\ 
& \hspace{1cm}  + \frac{q g^2 I_{avg}}{4} \Big( 2 \hat{\alpha}_{T}(0) + \frac{3\beta}{4}(\hat{\alpha}_{T}(f_0) + \hat{\alpha}_{T}(-f_0)) + \hat{\alpha}_{T}(2f_0) + \hat{\alpha}_{T}(-2f_0) + \frac{\beta}{4} (\hat{\alpha}_{T}(3f_0) + \hat{\alpha}_{T}(-3f_0))\Big) \label{eq:SImannex}
\end{align}

Equation~\ref{eq:SIm} then needs to be re-injected in equation~\ref{eq:SIm1} to take the time average.
It is expected that the terms $\frac{1}{T}\hat{\alpha}_{T}(f)$ tend to zero when T becomes large ($T \gg 1/f$).
The exception is when $f$ is a multiple of the laser repetition rate $f_r$.
In this case $\lim_{T\rightarrow \infty} \frac{1}{T}\hat{\alpha}_{T}(f) = 1$ (Annex~B).
For small values of the relative SRS modulation $\beta$, the terms containing $\beta$ in equation~\ref{eq:SIm} can be neglected, and the equation becomes:

\begin{equation}
S_{I_m}(f) = \frac{g^2 I_{avg}^2}{4} [S_{\alpha}(f-f_0) + S_{\alpha}(f+f_0)] + g^2 \frac{q I_{avg}}{2} + \frac{g^2}{4}[S_{\epsilon}(f-f_0) + S_{\epsilon}(f+f_0)] \label{eq:SIm2}
\end{equation}

Equation~\ref{eq:SIm2} is valid assuming no correlations between different frequencies in the laser intensity noise $\alpha$, or the electric noise $\epsilon$.
Additionally, the lock-in also applies a low pass filter, meaning the only relevant frequencies will be for $f < \Delta f \ll f_0$, where $\Delta f$ is the filter bandwidth defined in equation~\ref{eq:Bandwidth}.
One can expect $S_{I}$ to be slowly varying around $f_0$, and as a result this function will be approximated it by its value in $f_0$.
The single sided power spectral density for the mixed current then reads:
\begin{equation}
\label{eq:SIm4}
S^+_{I_m}(f) = \frac{g^2}{2} \Big( I_{avg}^2 S^+_{\alpha}(f_0) + 2 q I_{avg} + S^+_{\epsilon}(f_0)\Big) = \frac{g^2 I_{avg}^2}{2} \mathcal{RIN}_I(f_0, I_{avg})
\end{equation}

\section{Supplementary figures}

\clearpage

\bibliography{main}
\end{document}